\documentclass[aps,prb,amsfonts,amssymb,twocolumn,amsmath,preprintnumbers,nofootinbib,floatfix,superscriptaddress,
showpacs]{revtex4-1}
\usepackage[dvips]{graphics}
\usepackage{graphicx}
\usepackage{bm}

\begin{document}

\title{Bistable state in superconductor/ferromagnet heterostructures}
\author{I. V. Bobkova}
\affiliation{Institute of Solid State Physics RAS, Chernogolovka,
Moscow reg., 142432 Russia}
\affiliation{Moscow Institute of Physics and Technology, Dolgoprudny, 141700 Russia}
\author{A. M. Bobkov}
\affiliation{Institute of Solid State Physics RAS, Chernogolovka,
Moscow reg., 142432 Russia}

\date{\today}

\begin{abstract}
A novel state is predicted for a S/F heterostructure. It is shown that the equilibrium F/S bilayer can be in a bistable state: one state is ground and the other is metastable. One of these states is always normal and the other is superconducting. One can switch between these states by an external control parameter, for example, by an applied magnetic field. This phenomenon can manifest itself through the hysteresis behavior upon varying the applied magnetic field or, even, temperature. It can also lead to transition of the heterostructure into the superconducting state upon increase of the temperature.
\end{abstract}

\maketitle

Study of ferromagnet/superconductor (F/S) heterosructures is a hot topic in recent years. It is already well-known that they are much more than the sum of their parts. The proximity effect in such heterostructures is not reduced to the simple suppression of the superconductivity by the exchange field of the ferromagnet. A lot of nontrivial phenomena caused by the S/F proximity occur. Among them the oscillatory behavior of the Cooper pair wave function inside the ferromagnet \cite{buzdin05,golubov04}, which is known to cause $\pi$-Josephson junction formation \cite{buzdin82,ryazanov01} and the non-monotonic dependence of the critical temperature of S/F bilayers on the F layer thickness \cite{jiang95,zdravkov06,zdravkov10}. For inhomogeneous magnetization of the ferromagnet or in the presence of the spin-orbit coupling, the so-called long-range triplet superconducting correlations can occur \cite{bergeret05,robinson10,bergeret13}. The controllable long-range proximity effect can also be realized for homogeneous magnetization under nonequilibrium conditions \cite{bobkova12}. Also it is worth to note the possibility to realize the so-called in-plane FFLO-state in F/S bilayers \cite{mironov12} and to enhance its critical temperature by the orbital effect of the applied magnetic field \cite{bobkov13}.

In the present paper we predict another nontrivial phenomenon, which can take place in such heterostructures due to the proximity effect. The equilibrium F/S bilayer can be in a bistable state: one state is ground (corresponding to the global minimum of the free energy) and the other is metastable. One of these states is always normal and the other is superconducting. One can switch between these states by an external control parameter, for example, by an applied magnetic field. Which of the states (normal or superconducting) is ground and which is metastable depends on the particular parameters of the heterostructure and can be adjusted by the external control parameter. The bistability phenomenon can manifest itself through the hysteresis behavior upon varying the applied magnetic field or, even temperature.

Study of the bistability of superconductors in equilibrium and outside equilibrium \cite{galitskii84,elesin77} has a long history. Bistability in voltage-biased normal-metal/superconductor (N/S) heterostructures \cite{keizer06,snyman09,moor09} has been also investigated. The existence of two different states, which are stable at the same voltage is a common feature for nonequilibrium voltage-biased or current-biased S/N heterostructures. Here we predict that a similar phenomenon, two stable states, can occur in an equilibrium S/F heterostructure, which is not driven by an external current or voltage. While the possibility to have a bi-valued solution of the self-consistency equation was reported in the literature for F/S/F heterostructures \cite{tollis04}, the bistable state was not discussed yet. In general, systems possessing two stable states, which are divided by an energy gap, are of great interest because of possibility to use them as elements of quantum computation or as memory cells. An example of such systems on the basis of S/F/S heterostructures can be the so-called $\varphi$-Josephson junction \cite{mints98,buzdin03,goldobin11,sickinger12}, which has a doubly degenerate ground state with the Josephson phase $\pm \varphi$, where the value of $0< \varphi < \pi$ depends on design parameters. The effect predicted in the present work, can also be potentially of applied interest.

Now we proceed with the solution of the self-consistency equation for the superconducting order parameter $\Delta(T)$ for the S/F bilayer under applied magnetic field and demonstrate that it has two different solutions in a particular parameter range.  Let us assume that an external magnetic field $\bm H$ is applied in the plane of the bilayer [$yz$-plane]. We also choose the vector potential $\bm A=(0, H_z x, -H_y x)+\bm A_0 \equiv \bm A'x+\bm A_0$ to be parallel to the $yz$-plane. Here $\bm A' \equiv \bm H \times \bm {e_x}$. $\bm {e_x}$ is a unit vector along the $x$ direction, where $x$ is the coordinate normal to the S/F interface. In our calculations we assume that (i) S is a singlet s-wave superconductor; (ii) the system is in the dirty limit, so the quasiclassical Green's function obeys Usadel equations \cite{usadel}; (iii) the thicknesses of the S and F layers are assumed to be small with respect to the corresponding coherence lengths $d_{S(F)} \lesssim \xi_{S(F)}$. Here $\xi_S=\sqrt{D_S/\Delta_0}$ and $\xi_F=\sqrt{D_F/h}$ are the superconducting and magnetic coherence lengths, respectively; $D_{S(F)}$ is the diffusion constant in the superconductor (ferromagnet), $\Delta_0$ is the bulk value of the superconducting order parameter at zero temperature and $h$ is the exchange field of the ferromagnet. This condition allows us to neglect the variations of the superconducting order parameter and the Green's functions across the S and F layers.

The retarded Green's function $\check g^R \equiv \check g^R(\varepsilon,\bm r)$ depends on the quasiparticle energy $\varepsilon$ and the coordinate vector $\bm r=(x, \bm r_\parallel)$, where $\bm r_\parallel$ is parallel to the interface [$(yz)$ plane]. $\check ~$ means that the Green's function is a $4\times 4$ matrix in the direct product of spin and particle-hole spaces. We assume that the exchange field in the F layer is homogeneous $\bm h=(0,0,h)$. In this case there are only singlet and triplet with zero spin projection on the quantization axis pairs in the system. Then the normal $\hat g^R$ and anomalous $\hat f^R$ parts of the Green's function take the following structure in spin space: $\hat g^R(\varepsilon,\bm r)=[g_\uparrow (1+\sigma_3)/2+g_\downarrow (1-\sigma_3)/2]$ and $\hat f^R(\varepsilon,\bm r)=[f_\uparrow (1+\sigma_3)/2+f_\downarrow (1-\sigma_3)/2]i\sigma_2$, where $\sigma_{2,3}$ are the corresponding Pauli matrices in spin space. While we only consider the singlet pairing channel, the superconducting order parameter $\hat \Delta=\Delta i \sigma_2$. Further, it is convenient to exploit the well-known $\theta$-parametrization for the Green's function. Introducing the auxiliary quantity $\theta(\varepsilon,x)$ one can express the normal and anomalous parts of the Green's function as follows: $g_\sigma^R=-i\pi \cosh \theta_\sigma$ and $f_\sigma^R=i\pi \sinh \theta_\sigma e^{i \bm k \bm r_\parallel}$, where $\sigma=\uparrow,\downarrow$ and the phase factor $e^{i \bm k \bm r_\parallel}$ occurs if one consider the system under the applied magnetic field. In the framework of the $\theta$-parametrization Usadel equation takes the form:
\begin{eqnarray}
D \left[\partial_x^2 \theta_\sigma -(\bm k-\frac{2e}{c}\bm A)^2 \cosh \theta_\sigma \sinh \theta_\sigma \right]+ \nonumber \\
2 i (\varepsilon+\sigma h) \sinh \theta_\sigma -2i  \Delta \cosh \theta_\sigma = 0
\enspace .
\label{usadel}
\end{eqnarray}
Here $\sigma h=+h(-h)$ for $\theta_{\uparrow(\downarrow)}$. $D$ stands for the diffusion constant, which is equal to $D_{S(F)}$ in the superconductor (ferromagnet).

Eq.~(\ref{usadel}) should be supplied by the Kupriyanov-Lukichev boundary conditions \cite{kupriyanov88} at the S/F interface ($x=0$)
\begin{eqnarray}
\sigma_S \partial_x \theta_\sigma^S = \sigma_F \partial_x \theta_\sigma^F = g_{FS}\left.\sinh (\theta_\sigma^S - \theta_\sigma^F)\right|_{x=0}
\label{interface_cond}
\enspace ,
\end{eqnarray}
where $\sigma_{S(F)}$ stands for a conductivity of the S(F) layer and $g_{FS}$ is the conductance of the S/F interface. The boundary conditions at the ends of the bilayer are $\left. \partial_x \theta_\sigma^S \right|_{x=d_S} = \left. \partial_x \theta_\sigma^F \right|_{x=-d_F}=0 $.

\begin{figure}[!tbh]
  \centerline{\includegraphics[clip=true,width=3.0in]{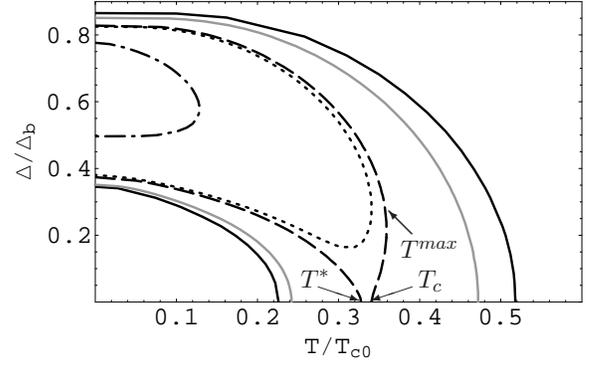}}
   \caption{Solution of the self-consistency equation $\Delta$ (in units of the zero-temperature bulk order parameter value $\Delta_b$) as a function of temperature (in units of the bulk critical temperature $T_{c0}$) for different magnetic fields. $H=0$ (black solid line); $0.015$ (gray solid); $0.0241$ (dashed); $0.025$ (dotted); $0.035$ (dashed-dotted). For the dashed line points $T^*$, $T_c$ and $T^{max}$ are shown. $H$ is measured in units of $\Phi_0/\pi\xi_F^2$. The other parameters of the system are $T_{c0}/h=0.1$, $D_Sh/(D_FT_{c0})=45$, $g_{FS}\xi_F/\sigma_S=10$, $\sigma_S/\sigma_F=6.0$, $d_F=0.5\xi_F$ and $d_S=4.2\xi_F$. }
\label{delta}
\end{figure}

Integrating Eq.~(\ref{usadel}) over $-d_F<x<0$ and $0<x<d_S$ and taking into account the boundary conditions, one can obtain the following  two coupled equations for $\theta_S$ and $\theta_F$ (which are assumed to be approximately constant in the S and F layers, respectively):
\begin{eqnarray}
\frac{iD_F}{2d_F}\left[ \frac{g_{FS}}{\sigma_F}\sinh(\theta_\sigma^F -\theta_\sigma^S)+\cosh \theta_\sigma^F \sinh \theta_\sigma^F \times \right.~~~~~~\nonumber \\
\left. \int \limits_{-d_F}^{0} dx \left(\bm k - \frac{2e}{c}\bm A(x)\right)^2 \right]+(\varepsilon+\sigma h)\sinh \theta_\sigma^F = 0
\label{theta1}
\enspace ,~~~~~~~~
\end{eqnarray}
\begin{eqnarray}
\frac{iD_S}{2d_S}\left[ \frac{g_{FS}}{\sigma_S}\sinh(\theta_\sigma^S -\theta_\sigma^F)+\cosh \theta_\sigma^S \sinh \theta_\sigma^S \times \right.~~~~~~\nonumber \\
\left. \int \limits_{0}^{d_S} dx \left(\bm k - \frac{2e}{c}\bm A(x)\right)^2  \right] + \varepsilon \sinh \theta_\sigma^S - \Delta \cosh \theta_\sigma^S = 0.~~
\label{theta2}
\end{eqnarray}
Solving Eqs.~(\ref{theta1})-(\ref{theta2}) together with the self-consistency equation
\begin{equation}
\Delta=\int \limits_{-\omega_D}^{\omega_D} \frac{d\varepsilon}{4} \Lambda \sum \limits_\sigma{\rm Re} \left[ \sinh \theta_\sigma^S \right] \tanh \frac{\varepsilon}{2T}
\enspace ,
\label{Tc}
\end{equation}
one obtains $\Delta(T,H,k)$ as a function of temperature, external magnetic field and wave vector $\bm k$. It is worth noting that $T$ and $H$ are external parameters, which can be controlled experimentally, but $\bm k$ is an intrinsic characteristic of the system and is determined from the energy consideration. Below we study solutions of Eq.~(\ref{Tc}) for an arbitrary $\bm k$. But in Fig.~\ref{delta} we plot $\Delta(T,H)$ as function of temperature for different magnetic fields and some particular value of $\bm k$, which corresponds to the extremum of $\Delta(T,H,k)$ with respect to $k$ for the parallel orientation of $\bm A$ and $\bm k$. For a given magnetic field all the curves, corresponding to all possible values of $k$, are located between the extremal solutions, shown in Fig.~\ref{delta}. 

From Fig.~\ref{delta} it is seen that at $H=0$ for $0<T<T_*(H=0)$ the self-consistency equation has two different nonzero solutions. It is worth noting that $\Delta=0$ is also a solution of the self-consistency equation. Here and below $T^*(H)$ stands for the lowest temperature, where the lower solution vanishes for a given applied field. $T^{max}(H)$ means the highest temperature, where the non-zero $\Delta$ exists for a given field.  $T_c(H)$ denotes the highest temperature, where the solution vanishes. For a dashed curve all these three temperatures are different and shown in Fig.~\ref{delta}. There is only rather narrow region of intermediate magnetic fields, where all these three temperatures are different. For lower magnetic fields $T_c(H)=T^{max}(H)$, as it is demonstrated by the black and gray solid curves. Upon increasing of the magnetic field $T^*(H)$ and $T_c(H)$ approach and finally merge at some particular field. For higher fields there is no temperature, where $\Delta$ vanishes, but the finite (not small) solutions of Eq.~(\ref{Tc}) still exist. That is, for high enough fields only $T^{max}$ exists. More high magnetic fields suppress superconductivity completely, as it should be. 

The both nonzero solutions of Eq.~(\ref{Tc}) for a given temperature (and the normal state $\Delta=0$) are extrema of the corresponding free energy functional. The higher $\Delta(T)$ is always minimum. That is, it represents the stable superconducting state of our bilayer. The lower $\Delta(T)$ always corresponds to the maximum of the free energy, that is absolutely unstable. It can be seen as follows. Let us consider the free energy of the system as a function of $\Delta$. In the vicinity of $\Delta=0$ we can restrict ourselves by the quadratic term $F^{(2)}=a\Delta^2$ of the GL free energy for the bilayer. The microscopic expression for $a$ can be obtained from the solution of the linearized (with respect to $\theta$) Usadel equation and takes the form
\begin{equation}
\left.a=N_F \left[ 1/\Lambda - \int \limits_{-\omega_D}^{\omega_D} \frac{d\varepsilon}{4\Delta} \sum \limits_\sigma{\rm Re} \left[ \theta_\sigma^S \right] \tanh \frac{\varepsilon}{2T} \right]\right|_{\Delta \to 0}
\label{a}
\enspace .
\end{equation}
Substituting $\theta_\sigma^S(\Delta \to 0)$, calculated from Eqs.~(\ref{theta1})-(\ref{theta2}) into Eq.~(\ref{a}), one obtains the following: (i) for all the magnetic fields, where the temperatures $T_c(H)$ and $T^*(H)$ exist, $a(T,H)>0$ for $T<T^*$ and $T>T_c$. This statement is valid for all possible values of $\bm k$. That is, the normal state is a minimum of the free energy for $T<T^*$, and, consequently, the lower nonzero $\Delta(T)$ corresponds to the maximum of the free energy. We have a bistable state, where the normal and the superconducting stable states are separated by an energy barrier. On the other hand, at $T^*<T<T_c$ the coefficient $a(T,H)<0$ at least for small values of $k$. Therefore, in the temperature range $T^*<T<T_c$ the normal state does not correspond to the minimum of the free energy. Now the only stable state is the superconducting state. For $T_c=T^{max}$ at temperatures $T>T_c$ only the normal state is stable. If the magnetic field is such that $T^{max} \neq T_c$ ($T^{max} > T_c$), then for $T_c<T<T^{max}$ the bistable state is again appears.  For high enough magnetic fields the temperatures $T_c(H)$ and $T^*(H)$ do not exist anymore and $a(T,H)$ is always positive. That is, the normal state always corresponds to the minimum of the free energy. Consequently, the lower $\Delta(T)$  is again represents the maximum of the free energy. So, the system is again in the bistable state at this temperature.

\begin{figure}[!tbh]
  \centerline{\includegraphics[clip=true,width=3.0in]{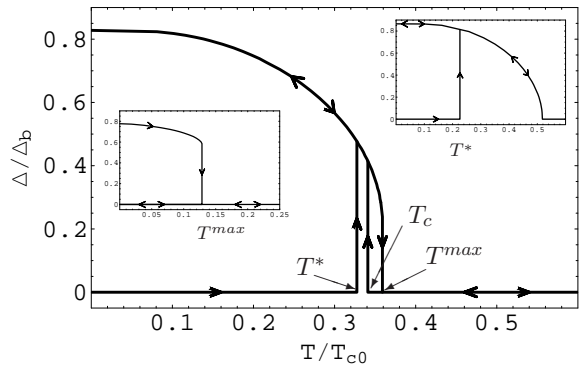}}
   \caption{Temperature evolution of the superconducting order parameter in the system. $H=0.0241$ for the main figure, $H=0$ for the right insert and $H=0.035$ for the left insert. The other parameters are the same as in Fig.~\ref{delta}.}
\label{delta_T}
\end{figure}

Now we turn to the discussion of possible experimental manifestations of this bistable state. We begin by considering the characteristic behavior of the system in the bistable state upon varying temperature at a given applied field. In dependence on the particular value of the field $H$ there are three different regimes of the temperature evolution of the superconducting order parameter $\Delta(T)$ in the system. They are demonstrated in Fig.~\ref{delta_T}. It is worth noting here that experimentally it is convenient to measure the electric resistance of the system (instead of $\Delta$) because we study the transitions between the superconducting and nonsuperconducting states of the system.

For low enough magnetic fields, which are determined implicitly by the condition $T_c(H)=T^{max}(H)$ (black and gray solid curves in Fig.~\ref{delta}), the temperature evolution of the system is represented in the right insert to Fig.~\ref{delta_T}. If the system is in the superconducting state at $T=0$ then it evolves gradually along the upper line upon increasing and decreasing temperature, as indicated by arrows and  undergoes transition to the normal state at $T=T_c(H)$. It can never get to the normal state at low temperatures. But if the system is already in the normal state at $T=0$, there is a transition to the superconducting state upon increasing temperature. This transition takes place at $T=T^*(H)$. How to make the system to get to the normal state at $T=0$ is discussed below. Upon further cooling and heating the system again cannot reach the normal state at $T=0$.

For high enough fields, determined by the condition that only $T^{max}(H)$ exists (dotted and dashed-dotted curves in Fig.~\ref{delta}), the situation is reverse. This is represented in the left insert to Fig.~\ref{delta_T}. Upon cooling the system evolves along the lower line. That is, it does not experience the transition to the superconducting state. If the system is in the superconducting state at $T=0$, then it gets to the normal state upon heating at $T=T^{max}(H)$ and never becomes superconducting upon further cooling.

There is a narrow region of intermediate magnetic fields, when $T^{max}(H)>T_c(H)$ (dashed line in Fig.~\ref{delta}), where the system can manifest a hysteresis behavior upon varying temperature. This case is represented in the main figure of Fig.~\ref{delta_T}.  Upon cooling or starting from the superconducting state at $T=0$ one can observe the hysteresis behavior, as indicated by arrows. The downward transition is at $T=T^{max}(H)$, and the upward transition corresponds to $T=T_c(H)$. However, if the system is in the normal state at $T=0$, upon heating it experiences the transition to superconducting state at $T=T^*(H)$ and after that upon further heating and cooling it cannot get to the normal state at $T=0$.

\begin{figure}[!tbh]
  \centerline{\includegraphics[clip=true,width=3.0in]{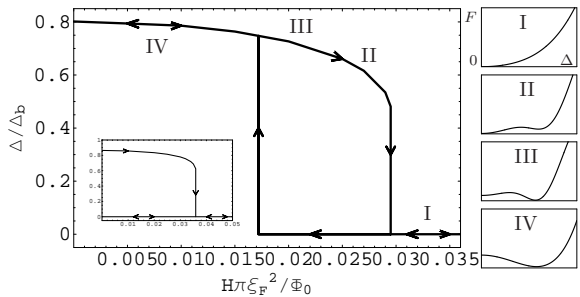}}
   \caption{Evolution of the superconducting order parameter in the system upon varying the magnetic field. Main figure: $T=0.25T_{c0}$, left insert: $T=0.1T_{c0}$. The right insert, consisting of four figures, represents the qualitative behavior (drawn by hand) of the free energy of the system as a function of $\Delta$ for different magnetic fields.}
\label{delta_H}
\end{figure}

Possibly, it is more convenient to study the bistable system in dependence on the applied field for a given temperature. Further we turn to discussion of this situation. If $T<T^*(H=0)$ then the system evolves as depicted in the insert to Fig.~\ref{delta_H} upon varying magnetic field. If the system is in the superconducting state at zero field, then upon increasing the field its behavior is standard: the field suppresses superconductivity. The transition takes place at the field determined implicitly by the condition $T=T^{max}(H)$. But starting from the normal state at high fields the system does not experience the transition to the superconducting state and keeps normal. This is the way to reach the normal state at zero temperature. Then one can observe the transition to the superconducting state upon increasing temperature, as discussed above. 

If the working temperature is such that there are two nonzero solutions of the self-consistency equation (See Fig.~\ref{delta}) for, at least, some values of the magnetic field, the system manifests the hysteresis behavior upon varying the magnetic field. This is demonstrated in the main figure of Fig.~\ref{delta_H}. The field of the upward transition is determined by $T=T^*(H)$ or by $T=T_c(H)$ (depending on the particular working temperature) and the field of the downward transition is determined by the condition $T=T^{max}(H)$. If the temperature is close enough to $T_c(H=0)$, the hysteresis behavior does not occur, because there is the only non-zero solution of the self-consistency equation for any magnetic field. 

One can also  discuss the hysteresis behavior by considering the free energy of the system. Its evolution with the field is depicted in the right inserts to Fig.~\ref{delta_H}. Starting from high fields the system evolves from the normal state (region marked as "I" in the figure, the only minimum of free energy at $\Delta=0$) to the superconducting state (region marked as "IV" in the figure, the only minimum of free energy at nonzero $\Delta$) via the bistable state (regions marked as "II" and "III" in the figure). Here the free energy has two minima - one for a normal state and the other for a superconducting state. At lower fields the global minimum corresponds to the superconducting state and at higher fields - to the normal state. Consequently, the thermodynamic transition between the superconducting and normal states should occur at some intermediate magnetic field. But we do not study the thermodynamic transition here because we do not consider the decaying rate of the metastable state due to fluctuations, this is beyond the scope of our mean field analysis.

\begin{figure}[!tbh]
   \begin{minipage}[b]{\linewidth}
     \centerline{\includegraphics[clip=true,width=2.2in]{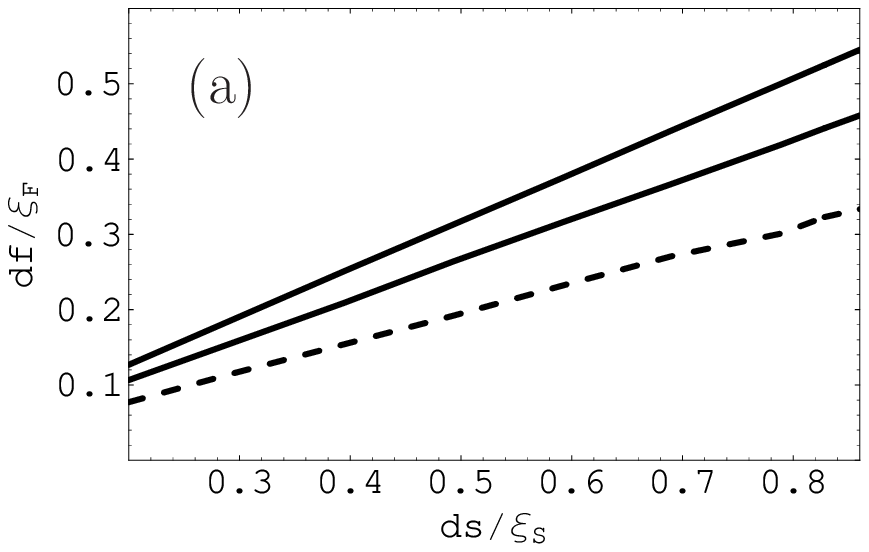}}
     \end{minipage}\hfill
    \begin{minipage}[b]{\linewidth}
   \centerline{\includegraphics[clip=true,width=2.2in]{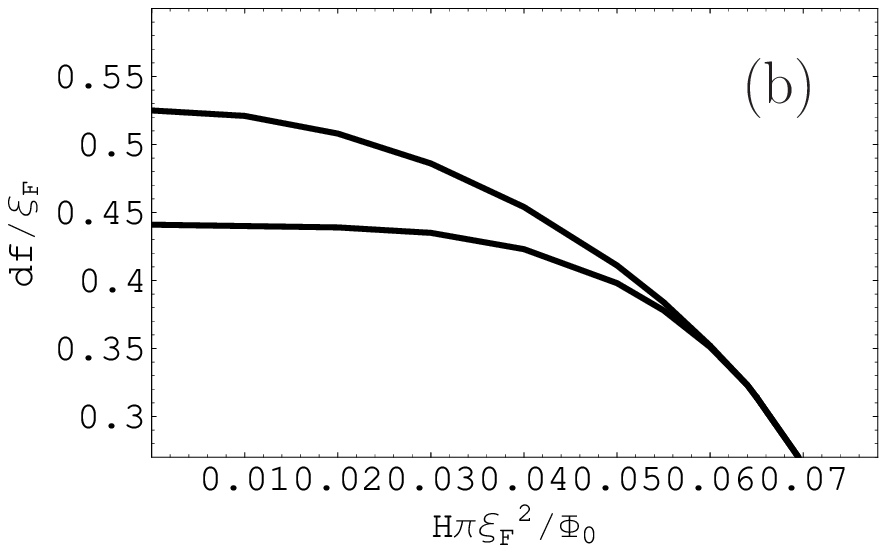}}
  \end{minipage}
   \caption{(a) Phase diagram of the S/F bilayer in the $(d_S,d_F)$-plane. The bistable state at $H=0$ is realized between the solid lines. At non-zero $H$ the bistable state can be realized between the upper solid and the dashed lines. (b) Phase diagram of the S/F bilayer in the $(H,d_F)$-plane. $d_S=4.2\xi_F=0.83\xi_S$. The bistable state is again between the solid lines. The other parameters of the bilayer for the both panels are the same as in Fig.~\ref{delta}.}
\label{phase}
\end{figure}

Now we discuss the conditions under which it is possible to realize the bistable state in the S/F bilayer. Fig.~\ref{phase}(a) represents the phase diagram of a S/F bilayer in the $(d_S,d_F)$-plane. The other parameters of the bilayer are taken to be close to the typical experimental values for a Nb/CuNi bilayer (see caption to Fig.\ref{delta} for the particular values). The solid lines give the boundaries of the normal, bistable and superconducting phases at $H=0$. The bistable state is between the solid lines. For $d_S \gtrsim \xi_S$ we are not able to conclude how the phase diagram evolves, because our calculation is restricted by the condition $d_S \lesssim \xi_S$.

It is seen that the bistable state can occur in the vicinity of the point, where the superconductivity is suppressed by the ferromagnet. In the region between the lower solid and the dashed lines the bistable state does not exist at zero applied field, but occurs at some finite external field. This is illustrated in Fig.~\ref{phase}(b).  Here the bistable state is again between the solid lines. The case, demonstrated in Fig.~\ref{phase} corresponds to the high enough (but not absolute) transparency of the S/F interface $g_{FS}\xi_F/\sigma_S=10$. Decrease of the interface transparency leads to the twofold consequences. From the one hand, superconductivity in the bilayer is fully suppressed at larger $d_F$. That is, the bistability region appears at larger $d_F$, what is more favorable for the experimental realization. From the other hand, the region of $d_F$, where the bistable state can exist (the distance between the solid lines in Fig.~\ref{phase}) shrinks. Therefore, it is desirable to have good enough interfaces in order to observe the bistable state.    

In conclusion, it is predicted that a S/F bilayer can possess a novel state. It is a bistable state - one of the states is always superconducting and the other is normal. One can switch between them by an external parameter. In the present paper we have considered an applied magnetic field as an external parameter. Which of the states is stable (superconducting or normal) and which is metastable, depends on the particular parameters of the system and can be easily adjusted by varying the control parameter. The bistable state can manifest itself experimentally by a hysteresis behavior in dependence on the magnetic field and temperature. Under appropriate conditions it can also lead to appearing of superconductivity upon temperature increase. Such a bistable state in S/F heterostructures can possibly be of an applied interest.                      




\end{document}